\documentclass[12pt]{article}
\usepackage{graphicx}

\newcommand\pubnumber{XXXX-yyy-zz}
\newcommand\pubdate{January 14, 2011}

\def\napoli{Purdue University\\
West Lafayette, Indiana 47907, USA}
\def\support{\footnote{Work supported by the U.S. Department of Energy Grant DE-FG02-91ER40681A29 to Purdue University.
}}

\def\Title#1{\begin{center} {\Large #1 } \end{center}}
\def\Author#1{\begin{center}{ \sc #1} \end{center}}
\def\Address#1{\begin{center}{ \it #1} \end{center}}

\newcommand\pubblock{\rightline{\begin{tabular}{l} \pubnumber\\
         \pubdate  \end{tabular}}}
\newenvironment{Abstract}{\begin{quotation}  }{\end{quotation}}
\newenvironment{Presented}{\begin{quotation} \begin{center}
             PRESENTED AT\end{center}\bigskip
      \begin{center}\begin{large}}{\end{large}\end{center} \end{quotation}}
\def\Acknowledgements{\bigskip  \bigskip \begin{center} \begin{large}
             \bf ACKNOWLEDGEMENTS \end{large}\end{center}}
\def\beq{\begin{equation}}
\def\eeq#1{\label{#1}\end{equation}}
\def\eeqn{\end{equation}}

\def\beqa{\begin{eqnarray}}
\def\eeqa#1{\label{#1}\end{eqnarray}}
\def\eeqan{\end{eqnarray}}

\let\bar=\overbar

\def\Dslash{\not{\hbox{\kern-4pt $D$}}}
\def\dslash{\not{\hbox{\kern-2pt $\del$}}}

\def\msb{{\bar{\ssstyle M \kern -1pt S}}}

\newcommand{\vcq}{V_{cd(s)}}

\newcommand{\ipb}{{\rm pb}^{-1}}

\newcommand{\bmath}{\begin{displaymath}}
\newcommand{\emath}{\end{displaymath}}

\newcommand{\fz}{f_+(0)}

\newcommand{\chargedpienu}{D^0\to \pi^- e^+ \nu_e}

\newcommand{\enu}{e^+ \nu_e}

\newcommand{\DD}{D\bar{D}}

\newcommand{\vcs}{|V_{cs}|}
\newcommand{\vcd}{|V_{cd}|}
\newcommand{\vub}{|V_{ub}|}

\newcommand{\beqn}{\begin{equation}}

\newcommand{\ra}{\rightarrow}

\def\cleoc{\hbox{CLEO-c}}

\begin{document}
\begin{titlepage}
\pubblock

\vfill
\Title{Charm Semileptonic Decays at CLEO-c}
\vfill
\Author{Bo Xin\support ~for the CLEO Collaboration}
\Address{\napoli}
\vfill
\begin{Abstract}
We review the recent results on $D$ and $D_s$ meson semileptonic decays from \cleoc.
Comparisons with lattice quantum chromodynamics (LQCD) calculations and
implications for $B$ physics are also discussed.
\end{Abstract}
\vfill
\begin{Presented}
The 6th International Workshop on the CKM Unitarity Triangle\\
University of Warwick, UK, September 6---10, 2010
\end{Presented}
\vfill
\end{titlepage}
\def\thefootnote{\fnsymbol{footnote}}
\setcounter{footnote}{0}

\section{Introduction}

In the Standard Model, the charge-changing transitions involving quarks
are described by the Cabibbo-Kobayashi-Maskawa~(CKM) matrix~\cite{ckm}.
Semileptonic decays are
the preferred way to determine the CKM matrix elements~\cite{richmanpole}. %, % due to its simplicity,
However, the power of semileptonic decays in probing the CKM matrix has been
severely limited by our knowledge of the strong interaction effects.
While techniques such as lattice quantum chromodynamics (LQCD)~\cite{fnallqcd, nad2k} offer increasingly precise calculations of the
hadronic form factors, experimental validation of these predictions is highly desired.
In charm semileptonic decays, the CKM matrix elements $\vcd$ and $\vcs$ are tightly constrained by CKM unitarity.
Therefore, precise measurements of charm semileptonic decay rates enable rigorous tests of
theoretical calculations of the form factors.
A validated theory can then be applied to the $B$ sector of flavor physics with increased confidence to determine $\vub$.

Studies of the exclusive semileptonic decays of the $D$ and $D_s$ mesons are also important for gaining a complete understanding
of charm semileptonic decays, and %have further implications such as those on $B_s \ra J/\psi f_0(980)$~\cite{stonejpsif0}.
as a probe of quark content and properties of the final state hadron.

\section{Experimental techniques}
\label{sec:tech}

In the past a few years, the experimental precision in charm semileptonic decays has been greatly improved.
At \cleoc, the dominant semileptonic %decay
analysis technique is $D$ tagging.
The $D$ mesons are produced through the decays $e^+e^- \ra \psi(3770) \ra \DD$ at the center-of-mass energy near 3.770~GeV.
This is a particularly clean environment since there is not enough energy to produce any additional particles other than the $\DD$.
The presence of two $D$ mesons in a $\psi(3770)$ event allows
a tag sample to be defined in which a $\bar{D}$
is reconstructed in a hadronic decay mode.
A sub-sample is then formed in which a positron and a set of hadrons, as a signature
of a semileptonic decay, are required in addition to the tag.
Tagging a $\bar{D}$ meson in a $\psi(3770)$ decay provides a $D$
with known four-momentum, allowing a semileptonic decay to be
reconstructed with no kinematic ambiguity, even though the
neutrino is undetected.

At $\sqrt{s}$=4.170 GeV,
the $D_s$ mesons are dominantly from $e^+e^-\ra D^*_s D_s$~\cite{dsscan}.
$D_s$ tag candidates are selected using hadronic final states,
then combined with well reconstructed photons. % to calculate the missing mass squared.
The four-momentum of the tag $D_s$ and photon combination yields the four-momentum of the signal $D_s$.

\section{Exclusive semileptonic decays of the $D$ mesons}

For pseudoscalar-to-pseudoscalar semileptonic decays, when the lepton mass is negligibly small,
the strong interaction dynamics can be described by a single form factor $f_+\left(q^2\right)$,
where $q^2$ is the invariant mass of the lepton-neutrino system.
The rate for a $D$ semileptonic decay
to a $\pi$ or $K$ meson
is given by
\begin{equation}
\frac{d\Gamma(D\rightarrow \pi(K) e\nu)}{dq^2}=X\frac{G_F^2\left|\vcq\right|^2}{24\pi^3}p^3\left|f_+\left(q^2\right)\right|^2,
\label{eq:diffrate}
\end{equation}
where $G_F$ is the Fermi constant, $\vcq$ is the relevant CKM matrix element,
$p$ is the momentum of the $\pi$ or $K$ meson in the rest frame of the parent $D$,
and $X$ = 1 or $1/2$ is a multiplicative factor due to isospin. %, equal to 1 for all modes except $\neutralpienu$, where it is $1/2$.

\begin{figure*}[bptb]
\centering
  \includegraphics*[width=4in]{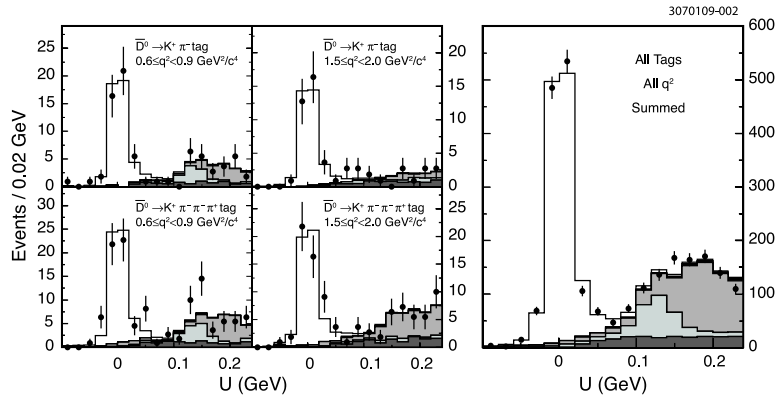}
  \caption{Fits of the $U$ distributions for $\chargedpienu$ (only a subset are shown).
  The unshaded histograms are signal.
  See Ref.~\cite{818kpienu} for details of the background components.}
  \label{fig:u_pi}
\end{figure*}

After a tag is identified, a positron and a set of hadrons are searched for in the recoiling system against the tag.
Semileptonic decays are identified using the variable $U \equiv
E_{\rm miss} - c|\vec{p}_{\rm miss}|$, where $E_{\rm miss}$ and
$\vec{p}_{\rm miss}$ are the missing energy and momentum of the
$D$ meson decaying semileptonically. %, calculated using the
Properly reconstructed decays are separated from backgrounds using an unbinned maximum-likelihood fit,
executed independently for each semileptonic mode, each tag mode, and each $q^2$ bin.
A sample of the $U$ distributions for $\chargedpienu$ is shown in Fig.~\ref{fig:u_pi}.
The signal and background shapes of the fits are taken from Monte Carlo samples.

The partial rates are then obtained by inverting the efficiency matrices, which account for both efficiency
and the smearing across $q^2$ bins.
Least squares fits are made to these partial rates, using several form factor parameterizations,
among which
the model-independent series expansion~\cite{srp} is generally of most interest.
Short surveys of these form factor parameterizations can be found in Refs.~\cite{818kpienu} and \cite{281kpienu} and references therein.

\begin{figure}[bptb]
\centering
  \includegraphics*[width=2.2in]{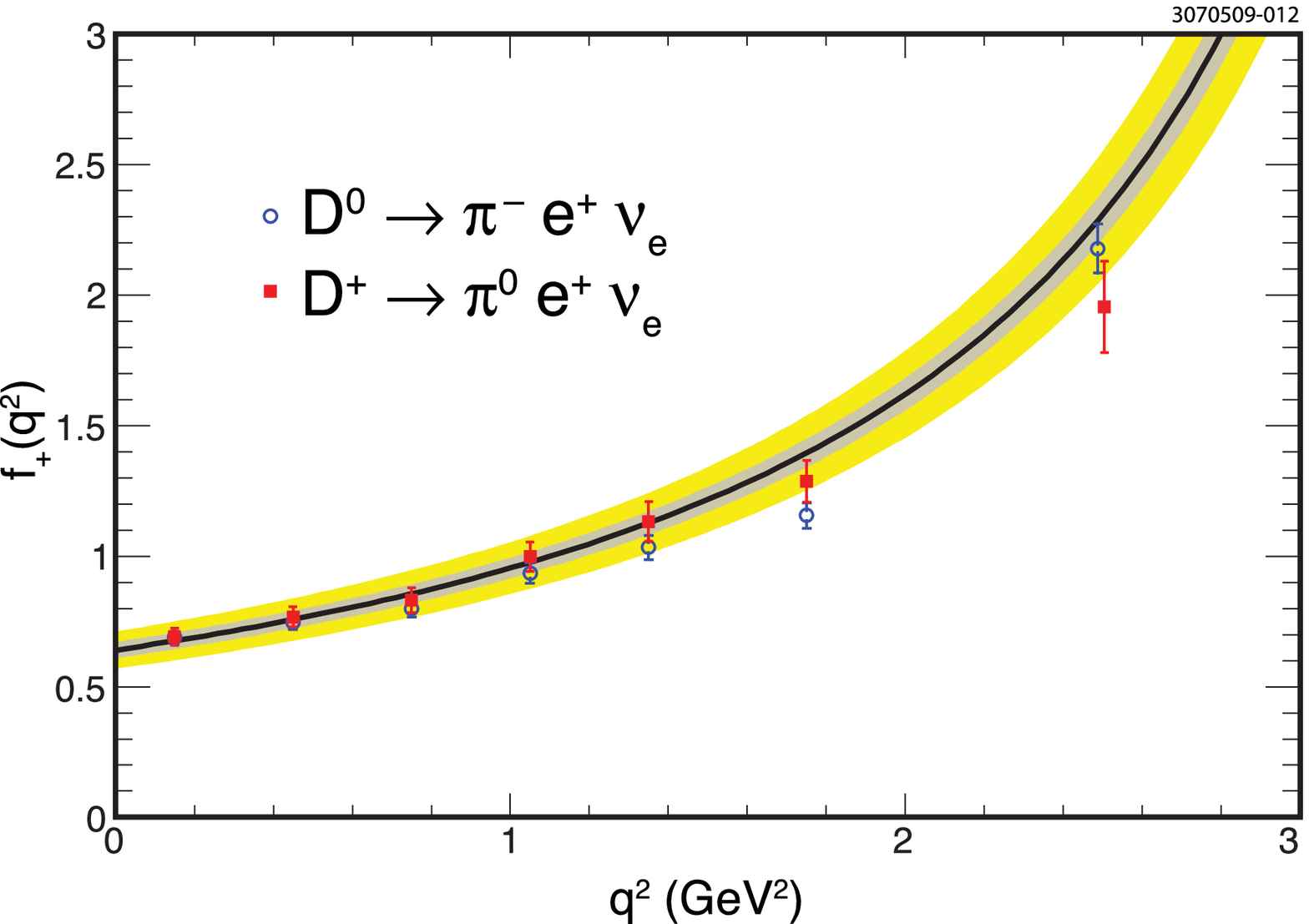}
  \includegraphics*[width=2.2in]{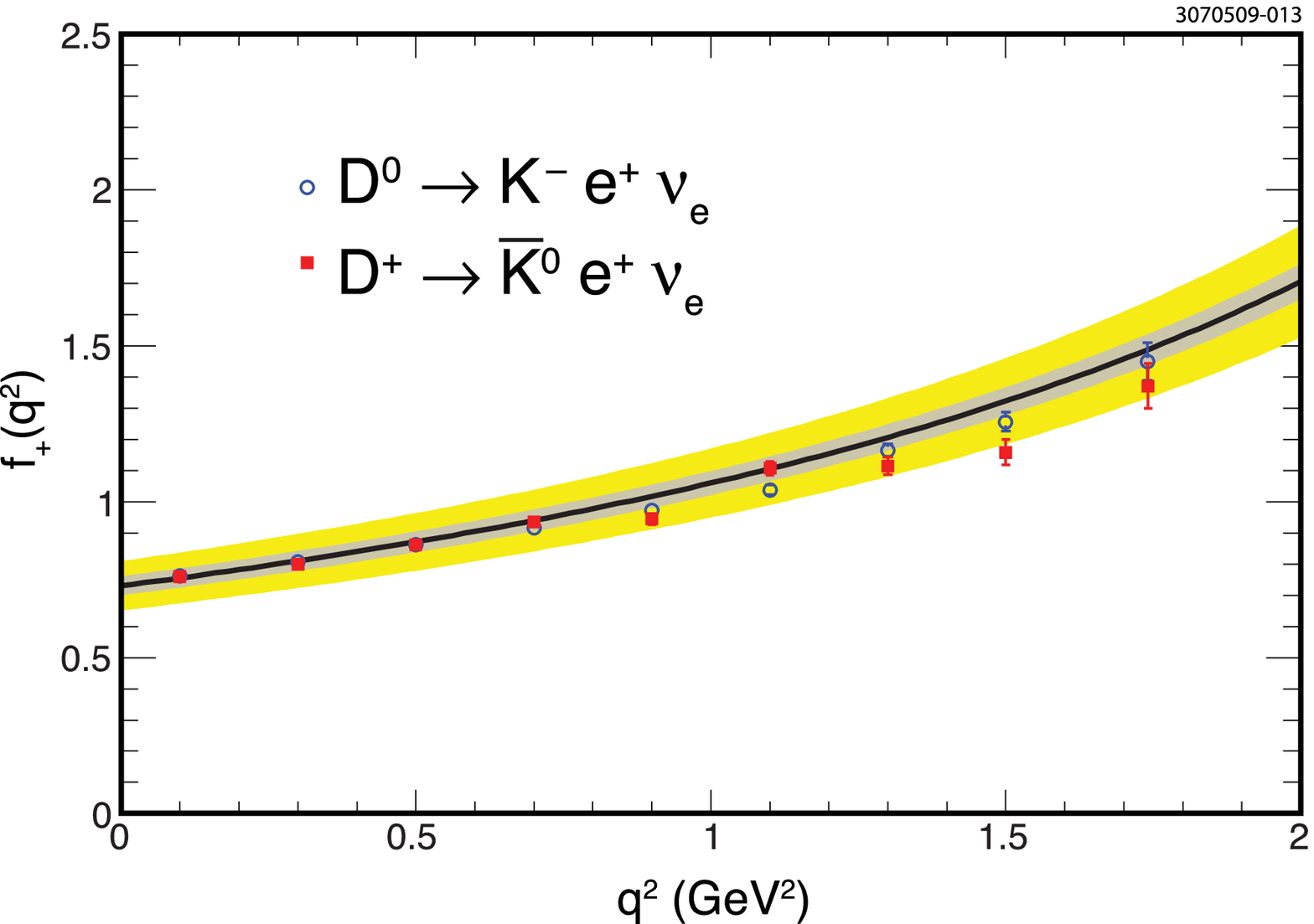}
  \caption{$f_+(q^2)$ comparison between isospin conjugate modes
  and with LQCD calculations~\cite{fnallqcd}.
    The solid lines represent LQCD fits to the modified pole model. %~\cite{modpole}.
    The inner bands show LQCD statistical uncertainties, and the outer bands the sum in quadrature of LQCD statistical and systematic uncertainties.}
  \label{fig:kpienuvslqcd}
\end{figure}

In Fig.~\ref{fig:kpienuvslqcd}, our Form factor shapes %using the 818 $\ipb$ of data
are compared between isospin conjugate modes and with the latest LQCD calculations~\cite{fnallqcd}.
Our results agree with LQCD calculations within uncertainties,
but are much more precise.
The LQCD bands are obtained using the modified pole model~\cite{modpole}.
The agreement between experiment and LQCD is better at low $q^2$ than high $q^2$.

Recently, a new HPQCD calculation~\cite{nad2k}, which employs a new approach for chiral/continuum extrapolations of $f_0(q^2)$ and uses the kinematic variable \lq\lq z" as in the series expansion, determines $f_+^{D\ra K}(0)$ with a theory error a factor of 4 smaller than previous unquenched lattice results~\cite{fnallqcd}.

Taking the $\left|\vcq\right|\fz$ values from the isospin-combined three parameter series expansion fits and using the latest LQCD measurements
for $f_+(0)$~\cite{nad2k}, we ~%extracts the magnitudes of CKM matrix elements $\vcd$ and $\vcs$, and
find
$\vcd = 0.234\pm0.007\pm0.002\pm0.025$ and $\vcs = 0.963\pm0.009\pm0.006\pm0.024$,
where the third uncertainties are from the LQCD calculation of $f_+(0)$.
These are in agreement with those based on the assumption of CKM unitarity~\cite{PDG2008}.
Our $\vcs$ measurement is the most precise direct determination. The $\vcd$ measurement is the most precise using semileptonic decays.

The form factors in $P \ra V$ transitions are studied at \cleoc.
The form factor measurement in $D\ra\rho\enu$ is the first form factor measurement in Cabibbo suppressed $P\ra V$ transitions.
When combined with the form factor in $D\ra K^* \enu$, the form factor in $D\ra\rho\enu$ helps in determining $|V_{\rm ub}|$
using the double ratio method~\cite{rhoenudoubleratio}.
%A total of $131\pm13$ ($173\pm15$) events are reconstructed in $D^0\ra \rho^-\enu$ ($D^+\ra \rho^0\enu$).
We find ${\mathcal B}(D^0\ra\rho^-\enu) = (1.77\pm0.11\pm0.10)\times 10^{-3}$
and ${\mathcal B}(D^+\ra\rho^0\enu) = (2.17\pm0.13\pm0.12)\times 10^{-3}$.
A four-dimensional log likelihood fit is performed to the isospin-conjugate modes simultaneously,
the form factor ratios~\cite{richmanpole} are found to be
$R_V = 1.48 \pm 0.15 \pm 0.05$ and $R_2 = 0.83 \pm 0.11 \pm 0.04$.

Using six hadronic tag modes,
we made a non-parametric form factor measurement in $D^+ \to K^-\pi^+ \enu$ and $K^-\pi^+ \mu^+ \nu_\mu$~\cite{kpilnu-non-para}.
%where the $\mu$ channel is studied along with the $e$ channel.
The $\mu$/$\pi$ separation is based on several cuts such as the invariant mass of the $K^-\pi^+$ candidate, and difference between $E_{\rm miss}$ and $|\vec{p}_{\rm miss}|$.
Muons enable the study of the mass-suppressed helicity form factor $H_t(q^2)$.
The form factor study includes the resonant and non-resonant $K^-\pi^+$.
The projective weighting technique is used to distinguish the helicity basis form factors based on their contributions to the decay angular distribution.
No evidence for d- or s-wave $K^-\pi^+$ component is seen.

In addition to studying the existing modes with unprecedented precision, \cleoc \linebreak has many results
from its searches for new semileptonic modes. One of the most recent results is the studies of
$D^+\ra \eta/\eta'/\phi \enu$~\cite{818etaenu}, in which
the decay $D^+\ra \eta'\enu$ is observed in two distinct analyses with statistical significance of 5.6 and 5.8 standard deviations, respectively.
These analyses also provide the first form factor measurement and an updated branching fraction for $D^+\ra \eta\enu$,
and an upper limit for $D^+\ra \phi\enu$ which is twice as restrictive as our previous limit~\cite{281etaenu}.
%These new modes are important for gaining a complete understanding of charm semileptonic decays.

\section{$D_s$ exclusive semileptonic decays}

The first absolute branching fraction measurements of the $D_s$ semileptonic decays
have been made by \cleoc~\cite{dsexclu} using 310 $\ipb$ of data at $\sqrt{s}$=4.170 GeV.
Via the tagged analysis technique,
six exclusive semileptonic modes
are searched for. %The missing-mass-squared ($MM^2$) distributions are shown in Fig.~\ref{fig:mm2}.
Among these, ${\mathcal B}(D_s^+\to K^0 e^+ \nu_e)=(0.37\pm 0.10 \pm 0.02)$\% and ${\cal B}(D_s^+\to K^{*0} e^+
\nu_e)=(0.18 \pm 0.07 \pm 0.01)$\% are the first measurements of Cabibbo suppressed exclusive $D_s$ semileptonic
decays. The measurement of
${\cal B}(D_s^+\to f_0 e^+ \nu_e) \times {\cal B}(f_0
\to \pi^+\pi^-) =(0.13\pm 0.04 \pm 0.01)$\% is the first
direct evidence of a semileptonic decay including a scalar meson
in the final state.
By searching for several additional hadronic final states with two charge tracks with or without a $\pi^0$,
we find no evidence of other $D_s$ semileptonic decays.

\section{Inclusive semileptonic decays of $D^0$, $D^+$ and $D_s$}

Using the full sample of open-charm data collected at \cleoc, the charm and charmed-strange
meson inclusive semileptonic branching fractions are obtained~\cite{818incl}.
Knowledge about exclusive semileptonic modes and form factor models are used to extrapolate the spectra below the 200 MeV momentum cutoff.
The ratios of the semileptonic decay widths are determined to be
$\Gamma^{\rm SL}_{D^+}$/$\Gamma^{\rm SL}_{D^0}$ = 0.985 $\pm$ 0.015 $\pm$ 0.024 and
$\Gamma^{\rm SL}_{D^+_s}$/$\Gamma^{\rm SL}_{D^0}$ = 0.828 $\pm$ 0.051 $\pm$ 0.025.
The former agrees with isospin symmetry.
The latter ratio shows that there is an indication of difference
between charm and charmed-strange meson semileptonic decay widths.

\section{Conclusions}
The \cleoc ~semileptonic program has been highly successful.
Most of \cleoc ~charm semileptonic results have been updated using full data sets.
Among the many interesting results,
$D \to K\enu$ and $\pi\enu$ form factors are in general agreement with LQCD.
However, LQCD precision lags.
Our $\vcs$ measurement is the most precise direct determination. The $\vcd$ measurement is the most precise using semileptonic decays.

\Acknowledgements

I thank the organizers for such a wonderful conference at the University of Warwick.
Valuable discussions with H. Na and Z. Liu on LQCD calculations are appreciated.
I. Shipsey is thanked for very helpful discussions and suggestions on this manuscript.

\end{document}